\documentstyle[prl,aps,12pt]{revtex}
\input{epsf.sty}

\def\b\mu{{\bf \mu}}

\def\half{\frac{1}{2}}

\begin{document}
\draft
\title{{\Large {\bf Analytic Properties of Thermal}}\\
{\Large {\bf Corrected Boson Propagators}}}
\author{H.C.G. Caldas$^{\dagger }$\thanks{%
e-mail address: hcaldas@funrei.br} and M. Hott$^{\ddag }$\thanks{
e-mail address: hott@feg.unesp.br}}
\address{${\dagger}$Departamento de Ci\^{e}ncias Naturais, DCNAT \\
Universidade Federal de S\~{a}o Jo\~{a}o del Rei, UFSJ \\
Pra\c{c}a Dom Helv\'{e}cio, 74. 36300-000, S\~{a}o Jo\~{a}o del Rei, MG, Brazil}
\address{${\ddag}$UNESP, Campus de Guaratinguet\'a,\\
PO. Box: 205. 12516-410, Guaratinguet\'{a}, SP, Brazil}
\date{November, 2002}
\maketitle

\begin{abstract}
We investigate the analytic properties of finite-temperature self-energies
of bosons interacting with fermions at one-loop order. A simple
boson-fermion model was chosen due to its
interesting features of having two distinct couplings of bosons with
fermions. This leads to a quite different analytic behavior of the bosons
self-energies as the external momentum $K^{\mu }=(k^{0},\vec{k})$ approaches
zero in the two possible limits. It is shown that the plasmon and Debye
masses are consistently obtained at the pole of the corrected propagator
even when the self-energy is analytic at the origin in the
frequency-momentum space.
\end{abstract}

\vspace{1cm}


%11.10.Wx Finite-temperature field theory 
%12.39.-X Phenomenological quark models
%52.60.+h Relativistic plasma

\vspace{0.5cm}

\vspace{2cm}

\newpage

\section{Introduction}

It is well known that the analytic properties of self-energies at finite
temperature are extremely important, since they are related to physical
processes such as the {\it pole physics} \footnote{%
the dispersion relation, the plasmon and Debye masses, the damping and decay
rates, etc.}. Nevertheless, they have attracted some attention only in the past
decade, as pointed out by Weldon in Ref. \cite{Weldon1}. The existence of a
unique limit as the external momentum $K_{\mu }\to 0$ has been admitted only
if the internal lines of the loop have propagators with different masses 
\cite{Das1}. We show by using a simple theory that: {\bf (i)} even if the
self-energy is analytic at the origin in the frequency-momentum space
(although the limits need not commute \cite{Das2}), it still leads to the
plasmon and Debye masses which arise from the consistent calculation at the
pole of the corrected propagators and {\bf (ii) }the analyticity, at the
origin, exhibited by one of the one-loop graphs is due to its very peculiar
dependence on the external momenta.

The paper is organized as follows. In Sec. II, we discuss the one-loop
self-energy of the bosons due their interactions with fermions. In Sec.
III, we study the pole physics which lead to the correct determination of
the plasmon and Debye bosons masses. In this section we also obtain the high
temperature limit for these masses. In Sec. IV we analyze the dispersion
relation which relates the real and imaginary parts of a one-loop
self-energy. We conclude in Sec. V.

\section{The Basic Interactions}

Let us consider the boson-fermion interaction (we have ignored, for simplicity, isospin) described by the Lagrangian
density which is part of the Gell-Mann and Levy model in its broken chiral
symmetry phase \cite{gell-mann}

\begin{equation}
{\cal L}(\bar{\psi},\psi ,\phi _{i})=\overline{\psi }\left[ i\gamma ^{\mu
}\partial _{\mu }-m-g(\sigma +i\pi \gamma ^{5}
)\right] \psi +{\cal L}_{0}(\phi _{i}),
\end{equation}
where $g$ is a nondimensional positive coupling constant and ${\cal L}
_{0}(\phi _{i})$ is the free Klein-Gordon Lagrangian for the bosons.

More recently the Gell-Mann and Levy model has been considered to obtain the
thermal masses of the fermions and bosons at high temperature by using a
modified self-consistent resummation (MSCR) \cite{heron1}, to study the
renormalization of the effective action at finite temperature \cite{heron2}
and to study the issue of analyticity of bubble diagrams \cite{hott}.

To one loop order, the zero temperature retarded self-energies for the pion and sigma fields in the Minkowski space read

\begin{equation}  \label{eq2}
\Pi (K) = i g^2 \int \frac{d^4P}{(2 \pi)^4} ~{\rm Tr} \left[ (a\gamma_5 -b) 
\frac{1}{\not{\! }P + \not{\! }K - m} (a\gamma_5 -b) \frac{1}{\not{\! }P -m}
\right],
\end{equation}
where for the pion one has $a=1$ and $b=0$ and for the sigma $a=0$ and $b=i$%
. As is well known, the $\gamma_5$ matrix will be responsible for a minus
sign in pion self-energy which will bring the differences between the
corrections for the two bosons. After performing the trace, this expression
gives

\begin{equation}  \label{eq3}
\Pi(K)= - 4 i g^2 \int \frac{d^4P}{ (2 \pi)^4 } \frac{ P^2 + P^{\mu}K_{\mu}
+c~m^2}{[(P+K)^2 - m^2][ P^2 - m^2]},
\end{equation}
where $c=-1$ for the pion and $c=1$ for the sigma. This allows us to write

\begin{equation}  \label{eq4}
\Pi(K)_{\sigma}=\Pi(K)_{\pi} + \tilde \Pi(k_0,k),
\end{equation}
where $k \equiv |\vec k|$ and

\begin{equation}  \label{eq4.1}
\Pi(K)_{\pi}= 4 g^2 [{\cal F}+{\cal G}],
\end{equation}

\begin{equation}  \label{eq4.2}
\tilde \Pi(k_0,k)= 4 g^2 {\cal H},
\end{equation}
with 
\begin{equation}  \label{eq4.3}
{\cal F}(m,T)= -i \int \frac{d^4P}{ (2 \pi)^4 } \frac{1}{[(P+K)^2 - m^2]},
\end{equation}

\begin{equation}
{\cal G}(m,T,K)= -i \int \frac{d^{4}P}{(2\pi )^{4}}\frac{P^{\mu }K_{\mu }}{%
[(P+K)^{2}-m^{2}][P^{2}-m^{2}]},  \label{eq4.4}
\end{equation}
and 
\begin{equation}
{\cal H}(m,T,k_{0},k)= -i \int \frac{d^{4}P}{(2\pi )^{4}}\frac{2m^{2}}{%
[(P+K)^{2}-m^{2}][P^{2}-m^{2}]}.  \label{eq4.5}
\end{equation}

Since we are interested in studying the thermal effects on the analytic
structure of the self-energy, we shall take only the nonzero temperature
parts of the integrals above. Applying the usual finite temperature
techniques in the imaginary-time formalism from Eqs. (\ref{eq4.3}) to (\ref{eq4.5}), 
we find the following expressions

\begin{equation}  \label{eq4.6}
{\cal F}_{\beta}(m,T)=2\int_{0}^{\infty} \frac{p^2~dp}{ (2 \pi)^2 } \frac{%
n_{\psi}(\omega)}{\omega},
\end{equation}
where $n_{\psi}$ is the fermion distribuction function $n_{\psi}(\omega)=%
\frac{1}{exp(\omega/T)+1},~\omega \equiv \sqrt{p^2 + m^2}$, and

\begin{equation}  \label{eq4.8}
{\cal G}_{\beta}(m,T,K)=K^2\int_{0}^{\infty} \frac{p^2~dp}{ (2 \pi)^2 } \frac{%
n_{\psi}(\omega)}{\omega}~\frac{1}{4pk}~\ln \left[\frac{(2pk+k^2-k_0^2)^2
-4k_0^2\omega^2}{(-2pk+k^2-k_0^2)^2 -4k_0^2\omega^2} \right],
\end{equation}

\begin{equation}  \label{eq4.9}
{\cal H}_{\beta}(m,T,k_0,k)=-2m^2\int_{0}^{\infty} \frac{p^2~dp}{ (2 \pi)^2 } \frac{%
n_{\psi}(\omega)}{\omega}~\frac{1}{2pk}~\ln \left[\frac{(2pk+k^2-k_0^2)^2
-4k_0^2\omega^2}{(-2pk+k^2-k_0^2)^2 -4k_0^2\omega^2} \right],
\end{equation}
where $K^2=K^{\mu} K_{\mu}=k_0^2-k^2$.

From here on, for the sake of simplicity of the notation, we drop the 
subscript $\beta$ in the self-energies, whose limits are

\begin{equation}  \label{eq5}
\Pi(k_0=0,k \to 0)_{\pi}=\Pi(k_0 \to 0,k=0)_{\pi}=4g^2{\cal F}(m,T),
\end{equation}

\begin{equation}  \label{eq6}
\Pi(k_0 \to 0,k=0)_{\sigma}={\cal F}(m,T)-8 g^2 m^2 \int_{0}^{\infty} \frac{%
p^2~dp}{ (2 \pi)^2 } \frac{n_{\psi}(\omega)}{\omega^3},
\end{equation}

\begin{equation}
\Pi (k_{0}=0,k\to 0)_{\sigma }={\cal F}(m,T)-8g^{2}m^{2}\int_{0}^{\infty }%
\frac{dp}{(2\pi )^{2}}\frac{n_{\psi }(\omega )}{\omega }  \label{eq7}
\end{equation}
which shows that the successive limits do not coincide at the origin of the
external four-momentum only in the sigma self-energy, as has already been
shown recently in a slightly different manner \cite{hott}.

\section{The Pole Physics}

Although the pion self-energy by itself is analytic at the origin of the
frequency-momentum space, it is the pole of the corrected propagator which
has physical relevance. In order to investigate the effect of the thermal
corrections due to the interactions with the fermions, let us consider the
zero-temperature bosons as massless. Below we calculate the effective boson 
masses induced by the thermal medium.

\subsection{The thermal corrected pion propagator}

The thermal corrected boson propagator is given by

\begin{equation}  
\label{eq8}
{\cal D}_{\sigma,\pi}(\omega_n,\vec k)^{-1}={\cal D}_{0 ~
\sigma,\pi}(\omega_n,\vec k)^{-1}+\Pi(\omega_n,\vec k)_{\sigma,\pi}=%
\omega^2_n+{\vec k}^2+ m_{\sigma, \pi}^2+\Pi(\omega_n,\vec k)_{\sigma,\pi},
\end{equation}
where ${\cal D}_{0 ~ \sigma,\pi}(\omega_n,\vec k)$ is the tree-level boson
propagator.

\begin{center}
{\bf {\it The pion plasmon mass:} }
\end{center}

%\newline

It is well known that particles immersed in a hot medium have their
properties modified. As they propagate in this plasma, they become dressed
by the interactions. Examples of immediate consequences are the appearance
of an effective thermal mass and the damping rate of collective excitations 
\cite{Bellac,heron3}. As we are considering massless bosons ($m_{\sigma, \pi}^2=0$), 
at the pole of the pion corrected propagator at zero momentum ($k=0$), we have

\begin{equation}
\label{eq9}
k_{0}^{2}=\Pi (k_{0},k=0)_{\pi }=4g^{2}{\cal F}(m,T)\left[ 1+\frac{k_{0}^{2}%
}{4\omega ^{2}-k_{0}^{2}}\right]. 
\end{equation}
Since the right-hand side (RHS) of Eq. (\ref{eq9}) has singularities, we write $k_{0,~\pi
}=M_{\pi }-i\gamma _{\pi }$, where $M_{\pi }$ and $\gamma _{\pi }$ are real.
Let us now define ${\rm I}_{\pi }=\Pi (k_{0,~\pi }=M_{\pi }-i\gamma _{\pi
},k=0)_{\pi }$. This allows us to get the leading contribution for the
plasmon thermal mass as well as the (weak) damping rate, $\gamma \ll M$,
respectively, of the pion:

\begin{equation}  
\label{eq10}
M_{\pi}^2={\cal P}~{\rm Re}~{\rm I}_{\pi}=4g^2{\cal F}(m,T),
\end{equation}

\begin{equation}
\label{eq11}
\gamma _{\pi }=-\frac{1}{2M_{\pi }}{\cal P}~{\rm Im}~{\rm I}_{\pi }=\frac{%
g^{2}}{4\pi }M_{\pi }\left( 1-\frac{4m^{2}}{M_{\pi }^{2}}\right)
^{1/2}~n_{\psi }(M_{\pi }/2)\to \frac{g^{2}}{8\pi }M_{\pi},  
\end{equation}
where ${\cal P}$ is the principal part of the integral. The arrow 
in Eq. (\ref{eq11}) refers to the limit of vanishing fermion mass.

\begin{center}
{\bf {\it The pion Debye mass:} }
\end{center}

%\newline

Another example of a fundamental property of a plasma is the Debye mass, $
M_D $, whose inverse is the screening length for electric fields in the
plasma \cite{Kapusta}. We adopt the definition of the Debye mass in terms of
the location of the pole in the static propagator for complex $k$. It was 
shown by Rebhan \cite{Rebhan} that, both for QED and QCD, this definition is 
the correct one \footnote{In QED this definition provides for an exponential 
decay of the screened Coulomb potential, whereas in QCD it gives a gauge-invariant 
result\cite{Eric}.}:

\begin{equation}  \label{eq12}
k_{\pi}^2=-\Pi(k_0=0, k_{\pi}^2=-M_{\pi,~D}^2)_{\pi}~\to~M_{\pi,~D}^2=4g^2%
{\cal F}(m,T) \left[1+\frac{M_{\pi,~D}^2}{4p^2} \right],
\end{equation}
where in the last equation, before the identification $k_{\pi}^2=-M_{%
\pi,~D}^2$, we have expanded $\Pi(k_0=0, k_{\pi})_{\pi}$ in the limit $%
k_{\pi} \to 0$. The solution of Eq. (\ref{eq12}) is straightforward: 
\begin{equation}  \label{eq12_1}
M_{\pi,~D}^2= 4g^2 \frac{{\cal F}(m,T)}{1-g^2 \tilde {{\cal F}}(m,T)},
\end{equation}
with 
\begin{equation}  \label{eq12_2}
\tilde {{\cal F}}(m,T)=2\int_{0}^{\infty} \frac{dp}{ (2 \pi)^2 } \frac{%
n_{\psi}(\omega)}{\omega}.
\end{equation}

\subsection{The thermal corrected sigma propagator}

\begin{center}
{\bf \it {The sigma plasmon mass:} }
\end{center}

In the pole of the thermal corrected sigma boson propagator at zero
(three-)momentum, we have

\begin{equation}  
\label{eq13}
k_0^2=\Pi(k_0, k=0)_{\sigma}=4g^2{\cal F}(m,T) \left[1+\frac{k_0^2-4m^2}{%
4\omega^2-k_0^2} \right].
\end{equation}
Repeating the same steps as before, for $k_{0,~\sigma}=M_{\sigma}-i\gamma_{%
\sigma}$ and ${\rm I}_{\sigma}=\Pi(k_{0,~\sigma}=M_{\sigma}-i\gamma_{%
\sigma},k=0)_{\sigma}$, one finds

\begin{equation}  
\label{eq14}
M_{\sigma}^2={\cal P}~{\rm Re}~{\rm I}_{\sigma}=4g^2{\cal F}(m,T),
\end{equation}

\begin{equation}  
\label{eq15}
\gamma_{\sigma}=-\frac{1}{2M_{\sigma}}{\cal P}~{\rm Im}~{\rm I}_{\sigma}= 
\frac{g^2}{4\pi}M_{\sigma} \left(1-\frac{4m^2}{M_{\sigma}^2}
\right)^{3/2}~n_{\psi}(M_{\sigma}/2) \to \frac{g^2}{8\pi}M_{\sigma},
\end{equation}
which gives the same results as for the pion case in the zero fermion mass
limit.

\begin{center}
{\bf {\it The sigma Debye mass:} }
\end{center}

We find next the sigma Debye mass in our ``plasma'' in the same way as we
did for the pion:

\begin{equation}  
\label{eq16}
k_{\sigma}^2 =-\Pi(k_0=0, k_{\sigma}^2=- M_{\sigma,~D}^2)_{\sigma}~\to~
M_{\sigma,~D}^2=4g^2{\cal F}(m,T) \left[1+\frac{M_{\sigma,~D}^2-4m^2}{4p^2}
\right],
\end{equation}
which gives 
\begin{equation}  
\label{eq17}
M_{\sigma,~D}^2= 4g^2 \frac{{\cal F}(m,T) -m^2 \tilde {{\cal F}}(m,T)}{1-g^2 
\tilde {{\cal F}}(m,T)} = M_{\pi,~D}^2 -4g^2m^2 \frac{\tilde {{\cal F}} (m,T)
}{ 1-g^2 \tilde {{\cal F}}(m,T)}.
\end{equation}

It should be pointed out that since we are considering vanishing zero temperature 
boson masses, resummation techniques are required in some order of the 
perturbative expansion to recover its validity. In the model and at the order 
we are studing, a resummation of one-loop boson diagrams (which could be interpreted as 
the replacement of the $m_{\sigma,\pi}^2$ in Eq.(\ref{eq8}) by effective boson thermal masses) 
would only shift the bosons plasmon and Debye masses, since these effective masses does not 
enter the (fermion) loop. Therefore, the qualitative results found here would be kept. 
However, Nieves and Pal \cite{Nieves} considered another model where the problem of 
the nonanalyticity of the one-loop self-energy disapeared when the calculation was 
carried out using improved propagatores for the particles that appear in the 
internal lines of the loop diagrams.

\subsection{High temperature limit for the pion and sigma plasmon and Debye
masses}

The high temperature limit of the functions $\tilde {{\cal F}}$ and ${\cal F}
$ are \cite{Jackiw}

\begin{eqnarray}
\label{eq17-1}
\tilde {{\cal F}}(m,T) & = & \frac{1}{2 {\pi}^2} \left( - \frac{1}{2} \ln 
\frac{\mu}{\pi} - \frac{1}{2} \gamma + \frac{1}{2} \sum_{n = 1}^\infty \frac{%
1}{n} \left[ \left( 1+ \frac{\mu^2}{4 \pi^2 n^2} \right)^{-1/2} - 1 \right]
\right)  \nonumber \\
& = & \frac{1}{2 {\pi}^2} \left(- \frac{1}{2} \ln \frac{\mu}{\pi} - \frac{1}{%
2} \gamma + \frac{\zeta (3)}{16\pi} \mu^2 - \frac{3 \zeta (5)}{64\pi^2}
\mu^4 + {\cal O} (\mu^6) \right), 
\end{eqnarray}
where $\mu=\frac{m}{T}$, $\gamma = 0.57721 \cdots $ is Euler's constant and
the numerical values of the $\zeta$ function at important points are $\zeta
(2) = \pi^2/6$, $\zeta (3) = 1.2020 \cdots $, $\zeta (4) = \pi^4/90$, $\zeta
(5) = 1.0369 \cdots $, and so on, whereas ${\cal F}$ is given by

\begin{eqnarray}  \label{eq17-2}
{\cal F}(m,T)= \frac{T^2}{2 {\pi}^2} \left( \frac{\pi^2}{12} + \frac{1}{4}
\mu^2 \ln \frac{\mu}{\pi} + \frac{1}{4} \left( - \frac{1}{2} + \gamma
\right) \mu^2 - \frac{\zeta (3)}{32\pi} \mu^4 + {\cal O} (\mu^6) \right).
\end{eqnarray}
So one sees that $\tilde {{\cal F}}$ is less relevant at high temperature
than ${\cal F}$. Thus, the pion and sigma, respectively, plasmon masses are 
\begin{equation}
\label{eq17-3}
M_{\pi}^2=\frac{g^2T^2}{6} + \frac{g^2T^2}{2 {\pi}^2} \left( - \half + \gamma + \ln \frac{\mu}{\pi} \right) \mu^2
+ {\cal O} (\mu^4)~\buildrel H.T. \over\longrightarrow ~ \frac{g^2T^2}{6},
\end{equation}

\begin{equation}
\label{eq17-4}
M_{\sigma}^2=\frac{g^2T^2}{6} + \frac{g^2T^2}{2 {\pi}^2} \left( - \half + \gamma + \ln \frac{\mu}{\pi} \right) \mu^2 
+ {\cal O} (\mu^4)~\buildrel H.T. \over\longrightarrow ~ \frac{g^2T^2}{6},
\end{equation}
where $H.T.$ denotes the dominant term in the high temperature limit (or,
equivalently, the zero fermion mass limit), while the Debye masses are
written as 
\begin{eqnarray}
\label{eq17-5}
M_{\pi ,~D}^{2} &=&4g^{2}{\cal F}(m,T)\left[ 1+g^{2}\tilde{{\cal F}}
(m,T)+(g^{2}\tilde{{\cal F}}(m,T))^{2}+{\cal O}(g^{6}\tilde{{\cal F}}
(m,T)^{3})\right]  \nonumber \\
&=& \frac{g^{2}T^{2}}{6}\left[ 1-\frac{g^{2}}{(2\pi )^{2}}\left( \ln \frac{%
\mu }{\pi }+\gamma \right) +\left( \frac{g^{2}}{(2\pi )^{2}}\right)
^{2}\left( \ln \frac{\mu }{\pi }+\gamma \right) ^{2}+\cdots \right], 
\end{eqnarray}

\begin{eqnarray}
\label{eq17-6}
M_{\sigma ,~D}^{2} &=&M_{\pi ,~D}^{2}-4g^{2}m^{2}\tilde{{\cal F}}
(m,T)[1-g^{2}\tilde{{\cal F}}(m,T)]^{-1} \nonumber \\
&=& M_{\pi ,~D}^{2}+  \frac{g^{2}m^{2}}{\pi ^{2}}\left( \ln \frac{\mu }{\pi }%
+\gamma \right) \left[ 1-\frac{g^{2}}{(2\pi )^{2}}\left( \ln \frac{\mu }{\pi 
}+\gamma \right) +\left( \frac{g^{2}}{(2\pi )^{2}}\right) ^{2}\left( \ln 
\frac{\mu }{\pi }+\gamma \right) ^{2}+\cdots \right].
\end{eqnarray}

These results clearly show that, despite the pion self-energy being analytic
at the origin in the momentum space, its physical plasmon and Debye masses
are different, as they should be, thanks to the consistency of the calculations
at the pole of the corrected propagator.

\section{The dispersion relation}

Usually, the noncommuting limits have been traced back to the cut structure of
the one-loop self-energy through the dispersion relation \cite{Weldon2,Das1}

\begin{equation}  
\label{eq18}
{\rm Re}~ \Pi(k_0,k)=\frac{1}{\pi}{\cal P}~\int_{-\infty}^{\infty}du \frac{{\rm Im}%
~\Pi(u,k)}{u-k_0}.
\end{equation}

However this relation is not general. If $\Pi (k_{0},k)\sim
\,k_{0}^{n}$ $(n\geq 0)$ for $k_{0} \rightarrow \infty $, as is the case of
the pion self-energy in the approximation used here, a more general relation
should be used \cite{bogoliu,Kaku,Gross}. Particularly for the pion case we
have

\begin{equation}
{\rm Re}~\Pi (k_{0},k)={\rm Re}~\Pi (0,0)+\frac{1}{\pi }~(k_{0}^{2}-k^{2})
~{\cal P}~\int_{-\infty }^{\infty }du\frac{{\rm Im}~\Pi (u,k)}{u^{2}(u-k_{0})},
\label{disprel}
\end{equation}

\noindent 
with ${\rm Im}~\Pi (k_{0},k)$ given by

\begin{eqnarray}
{\rm Im}~\Pi (k_{0},k) &=&-\frac{1}{2}\int \frac{d^{3}\vec{p}}{(2\pi )^{2}}\,
\frac{1}{2\omega \Omega }\left\{ (\Omega +\omega )^{2}\left[ \delta
(k_{0}+\Omega +\omega )-\delta (k_{0}-\Omega -\omega )\right] \right.  \\
&&+\left. (\Omega -\omega )^{2}\left[ \delta (k_{0}+\Omega -\omega )-\delta
(k_{0}-\Omega +\omega )\right] \right\} \tanh\left(\frac{\beta \omega}{2} \right) \,,  
\nonumber
\end{eqnarray}

\noindent where $~\Omega \equiv \sqrt{(p-k)^{2}+m^{2}}$.

One can check that Eq. (\ref{disprel}) can also be used to show
that the analytic behavior of ${\rm Re}~\Pi (k_{0},k)$ is due to the
kinectic term multiplying the integral and that in both limits $
(k_{0}=0,k\rightarrow 0$ and $k=0,k_{0}\rightarrow 0)$ this contribution
vanishes leaving the first term as the sole contribution.

\section{Conclusions}

\label{conc} In this paper we have studied the analytic properties of
thermal corrected boson propagators through their interaction with fermions.
A particular model was chosen based on the fact that its two kinds of bosons
couple differently with the fermions, which leads to distinct (unexpected)
behavior of their self-energy. We have shown that the pion self-energy is
analytic at the origin in the frequency-momentum space at finite
temperature, whereas the sigma self-energy is nonanalytic. In spite of
this, we have shown that the analytic behavior found for the pion
self-energy does not spoil the difference between the plasmon and Debye
pion masses. We have also shown that the two physical masses arise from the
consistent calculation at the pole of the corrected propagator. Then, we
have derived the plasmon and Debye masses for both the pion and sigma
bosons. We note here that the answer to the question of which mass of a
certain field is manifested in a plasma in a given temperature $T$ depends
strictly on the situation encountered (or assumed) by its four momentum.
Besides, the specific dependence on the external momenta of the pion 
self-energy graph implies a modification in the usual dispersion relation 
which allows us to trace back the origin of the analyticity. Similar behavior 
seems also to be found in derivative coupling models \cite{Hott}. This has 
been shown to be a criteria other than the existence of distinct masses running in 
the upper and lower internal lines of a diagram \cite{Das1}, which could be 
used to predict the analytic behaviors of bubble diagrams as the external
momentum $K_{\mu }\to 0$.

\section*{Acknowledgments}

The authors would like to thank Dr. A.L. Mota for useful discussions.

%%%%%%%%%%%%%%%%%%%%%%%%%%%%%%%%%%%%%%%%%%%%%%%%%%%%%%%%%%%%%%%%%


\begin{references}


\bibitem{Weldon1}  H. Arthur Weldon, {Phys. Rev. }{\bf D 65}, 076010 (2002).

\bibitem{Das1}  P. Arnold, S. Vokos, P. Bedaque, and A. Das, {Phys. Rev. }
{\bf D 47}, 4698 (1993).

\bibitem{Das2}  A. Das, {\it Finite-Temperature Field Theory} (World
Scientific, New York, 1997).

\bibitem{gell-mann}  M. Gell-Mann and M. Levy, Nuovo Cimento {\bf 16}, 705
(1960).

\bibitem{heron1} H.C.G. Caldas, A.L. Mota, and M.C. Nemes, Phys. Rev. {\bf D
63}, 56011 (2001).

\bibitem{heron2}  H.C.G. Caldas and M.C. Nemes, Phys. Lett. B {\bf 523}, 293
(2001).

\bibitem{hott}  M. Hott and G. Metikas, Phys. Rev. {\bf D 65}, 085043 (2002).

\bibitem{Bellac}  Michel Le Bellac, {\it Thermal Field Theory} (Cambridge
University Press, Cambridge, England, 1996).

\bibitem{heron3} H.C.G. Caldas, Nucl.Phys.B {\bf \ 623}, 503 (2002), and references therein.

\bibitem{Kapusta}  J. Kapusta, {\it Finite-Temperature Field Theory}
(Cambridge University Press, Cambridge, 1989).

\bibitem{Rebhan}  A.K. Rebhan, {Phys. Rev. }{\bf D 48}, 3967 (1993).

\bibitem{Eric}  Eric Braaten and Agustin Nieto, {Phys. Rev. Lett. }{\bf 73},
2402 (1994).

\bibitem{Nieves}  J.F. Nieves and P.B. Pal, {Phys. Rev. }{\bf D 51}, 5300 (1995).

\bibitem{Jackiw}  L. Dolan and R. Jackiw, Phys. Rev. {\bf D} {\bf 9}, 3320
(1974), Appendix C.

\bibitem{Weldon2}  H. Arthur Weldon, {Phys. Rev. }{\bf D 47}, 594 (1993).

\bibitem{bogoliu}  N.N. Bogoliubov and D.V. Shirkov, {\it Introduction to
the Theory of Quantized Fields} (Interscience Publishers Ltd., London, 1959).

\bibitem{Kaku}  M. Kaku, {\it Quantum Field Theory: a Modern Introduction}
(Oxford University Press, Oxford, 1993).

\bibitem{Gross}  Franz Gross, {\it Relativistic Quantum Mechanics and Field Theory} (Wiley, New York, 1999).

\bibitem{Hott} H.C.G. Caldas and M. Hott (work in progress).

\end{references}
\end{document}